\documentclass[reprint,nofootinbib,superscriptaddress,amsmath,amssymb,aps,pra,floatfix]{revtex4-2}
\usepackage{graphicx,color}
\usepackage{bm}
\usepackage{hyperref}
\usepackage[normalem]{ulem}
\usepackage{braket}
\usepackage{placeins}

\def\vec#1{\bm{#1}}
\def\abs#1{\left\lvert#1\right\rvert}

\def\eb{\varepsilon_B}

\newcommand{\veck}{\mathbf k}

\newcommand{\vecp}{\mathbf p}
\newcommand{\vecq}{\mathbf q}

\newcommand{\ded}{\hat d^\dagger}
\newcommand{\de}{\hat d}
\newcommand{\bed}{\hat b^\dagger}
\newcommand{\be}{\hat b}

\newcommand{\ad}{\hat a^\dagger}
\renewcommand{\a}{\hat a}

\begin{document}
\title{Self-stabilized Bose polarons}
\author{Richard Schmidt}
\affiliation{Max-Planck-Institute of Quantum Optics,
	Hans-Kopfermann-Stra{\ss}e 1, 85748 Garching, Germany}
\affiliation{Munich Center for Quantum Science and Technology,
	Schellingstra{\ss}e 4, 80799 Munich, Germany
}
\author{Tilman Enss}
\affiliation{Institut f\"ur Theoretische Physik,
  Universit\"at Heidelberg, 69120 Heidelberg, Germany}
\date{\today}

\begin{abstract}
  The mobile impurity in a Bose-Einstein condensate (BEC) is a
  paradigmatic many-body problem.  For weak interaction between the
  impurity and the BEC, the impurity deforms the BEC only slightly and
  it is well described within the Fr\"ohlich model and the Bogoliubov
  approximation.  For strong local attraction this standard approach,
  however, fails to balance the local attraction with the weak
  repulsion between the BEC particles and predicts an instability
  where an infinite number of bosons is attracted toward the impurity.
  Here we present a solution of the Bose polaron problem beyond the
  Bogoliubov approximation which includes the \emph{local} repulsion
  between bosons and thereby stabilizes the Bose polaron even near and
  beyond the scattering resonance.  We show that the Bose polaron
  energy remains bounded from below across the resonance and the size
  of the polaron dressing cloud stays finite.  Our results demonstrate
  how the dressing cloud replaces the attractive impurity potential
  with an effective many-body potential that excludes binding.  We
  find that at resonance, including the effects of boson repulsion,
  the polaron energy depends universally on the effective
  range. Moreover, while the impurity contact is strongly peaked at
  positive scattering length, it remains always finite. Our solution
  highlights how Bose polarons are self-stabilized by repulsion,
  providing a mechanism to understand quench dynamics and
  nonequilibrium time evolution at strong coupling.
\end{abstract}
\maketitle


\section{Introduction}
\label{sec:intro}

Impurities in a Bose-Einstein condensate (BEC) exhibit a multitude of
fundamental physical phenomena: the formation of quasiparticles
\cite{devreese2009}, Efimov bound states \cite{zinner2013,
  levinsen2015}, synthetic Lamb shifts
\cite{rentrop2016,lemeshko2015}, Casimir interactions induced by a
fluctuating medium \cite{naidon2018, camacho2018bipolarons}, and
quantum criticality \cite{yan2020bose}.  Current experiments with
ultracold atomic gases are investigating several of these effects
reaching far into the strong-coupling regime \cite{catani2012,
  scelle2013, rentrop2016, hu2016, jorgensen2016, yan2020bose,
  skou2021}.  For understanding experimental observations it is thus
vital to develop a theoretical model that applies to impurity systems
at strong coupling and that can address both ground state and
non-equilibrium phenomena.

\begin{figure}[b]
  \centering
  \includegraphics[width=.9\linewidth]{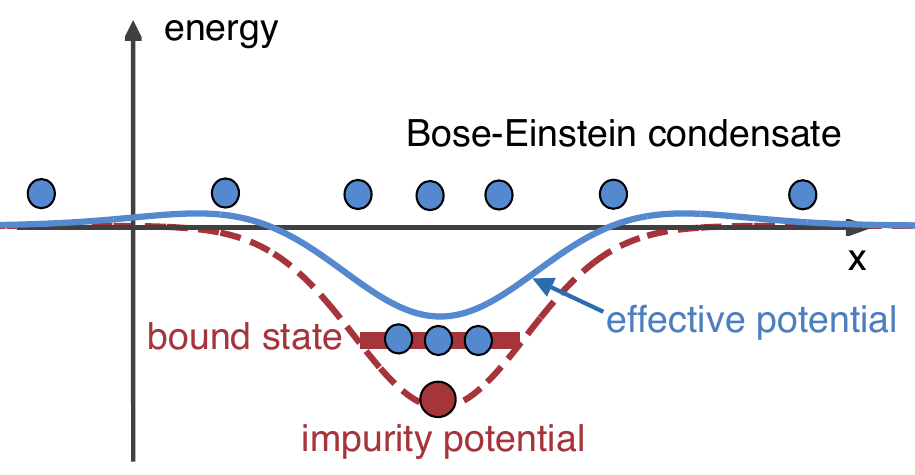}
  \caption{Illustration of the self-stabilized Bose polaron. The
    strong-coupling Bose polaron mimics a microtrap (red dashed line)
    with bound state (red bar) within the Bose-Einstein condensate.
    If several bosons (blue dots) occupy the bound state, boson
    repulsion results in a shallower effective potential seen by the
    remaining bosons (blue solid line) that no longer admits a bound
    state.  The dressing cloud itself thus stabilizes the Bose
    polaron.}
  \label{fig:microtrap}
\end{figure}

At weak interactions between the impurity and the BEC, the impurity
deforms the BEC only slightly and physics is well described within the
Fr\"ohlich model in terms of long-wavelength phonon excitations
\cite{devreese2009, froehlich1954, girardeau1961, astrakharchik2004,
  cucchietti2006, bruderer2008, tempere2009, casteels2011, shashi2014,
  grusdt2015renormalization, vlietinck2015, nielsen2019}.  For strong
local attraction, instead, the Fr\"ohlich model is incomplete and
needs to be amended by quadratic terms that absorb and re-emit phonons
\cite{rath2013}. Importantly, these term are also required to
correctly capture the formation of bound states between the impurity
and bath atoms which is a crucial ingredient for the physics of
polorans at strong coupling \cite{schmidt2018}.  Variational wave
functions based on a single phonon excitation \cite{chevy2006, li2014}
are able to describe the single occupation of such a bound state.
However, the bosons that make up the BEC tend to bunch, and at strong
coupling it is energetically favorable to occupy the bound state
multiple times leading to a gain of several times the binding energy.
This process, recently observed with Rydberg atoms immersed in a BEC
\cite{Camargo2018}, is described by a coherent state variational
ansatz \cite{rath2013, shchadilova2016quantum, grusdt2017,schmidt2018}
that allows for an arbitrary number of excitations and strong local
deformations of the condensate \cite{Massignan2005b,
  Astrakharchik2015, ashida2018, chen2018, drescher2019,
  hryhorchak2020, drescher2021}.

Generally, the application of the variational principle for the
determination of the ground state is only viable if the Hamiltonian is
bounded from below. Under this condition a stable solution can be
found, and theoretical approaches should rely only on such
approximations that preserve stability of the underlying
problem\footnote{In this work we restrict ourselves to models of cold
  dilute gases that disregard deeply bound states, transitions to
  solid or liquid phases, as well as large cluster formation.}. For
the strongly interacting Bose polaron, the resulting theoretical
challenge can be understood in the simple toy model illustrated in
Fig.~\ref{fig:microtrap}.  Here a static attractive impurity is
represented by a local potential well of finite range around the
impurity.  This well acts as a microtrap within the BEC.  At strong
coupling beyond a scattering resonance \cite{chin2010feshbach}, the
potential well (red dashed line) is deep enough to admit a bound state
with energy $\varepsilon_i=-\eb<0$ (red bar).  In absence of boson
repulsion, in the many-body ground state \emph{all} bosons would
occupy the bound state and the ground-state energy $E_0=-N\eb$ is
indeed unbounded from below in the thermodynamic limit: the whole BEC
is collapsed onto the impurity.

Naturally, a local boson repulsion counteracts this process by
balancing the impurity attraction \cite{drescher2020} and thus
providing a lower bound to the ground-state energy.  This mechanism
can be understood in terms of a single-site Bose-Hubbard model
\cite{bloch2008, massignan2021} with local repulsion
$\frac U2n_i(n_i-1)$ on the impurity site $r_i$, that competes with a
local attractive potential energy $-\eb n_i$ for occupation number
$n_i$.  For $U>0$ the ground state has a finite occupation
$n_i\simeq 2\eb/U$, which is nonperturbative in the strength of
the interaction: the size of this polaron dressing cloud grows for
weaker repulsion. It is crucial to adequately capture this repulsive
effect in the theoretical description of Bose polarons.

Previous variational approaches have included the boson repulsion only
at the level of the Bogoliubov approximation.  Here the interaction
between the Bogoliubov quasiparticles is neglected, and thus bosons in
the bound state fail to generate the compensating pressure required to
ensure the stability of the ground state. As a consequence this
approach falsely predicts instead a dynamical instability in presence
of boson repulsion toward infinite occupation of the bound state in
the strong coupling regime \cite{shchadilova2016quantum, grusdt2017,
  drescher2019}.  This shows the crucial importance of including the
boson repulsion beyond the Bogoliubov approximation.

In this work, we present a stable variational approach to the Bose
polaron problem at strong coupling. Our approach applies to arbitrary
dimension and impurity-boson scattering lengths and it provides a
basis for the study of dynamical properties of Bose polarons. In this
way our work complements and extends previous approaches using quantum
Monte Carlo to determine ground-state properties
\cite{penaardila2015}, studies of the role of boson repulsion in
one-dimensional systems \cite{Dehkharghani2015, Mistakidis2019b,
  mistakidis2019, jager2020, mistakidis2020}, or a recently developed
nonlocal Gross-Pitaevskii (GP) theory for nonequilibrium dynamics
\cite{drescher2020}. Moreover, previous works raised the question to
which extent polaron properties are universal in the short-range limit
\cite{yoshida2018, field2020}. In the following we study this question
across the Feshbach resonance, and in particular in the regime where a
bound state is supported by the impurity-bath potential, extending
previous work that considered how the properties of the Bose polaron
depend on the range of the interactions, both for the finite-range
boson repulsion in the nonlocal GP theory \cite{drescher2020} and for
finite-range impurity potentials \cite{guenther2021, massignan2021,
  astrakharchik2021}.

Specifically, we study the effect of local boson repulsion and
finite-range attractive impurity potentials employing an inhomogeneous
variational state that allows for large dressing clouds and strong
local deformations of the BEC. In Sec.~\ref{sec:model} we introduce
the stable Bose polaron model and discuss its solution within
Gross-Pitaevskii theory.  In Section~\ref{sec:effpot} we minimize the
resulting GP energy functional and obtain the condensate wave function
around the impurity. We find that even if the bare impurity potential
admits a bound state, the emerging effective potential does not, thus
providing a simple mechanism for the self-stabilization of Bose
polarons.  Section~\ref{sec:energy} presents our results for the
polaron energy, the size of the polaron dressing cloud and the Tan
contact across the resonance.  The universality of the polaron energy
is discussed in Sec.~\ref{sec:univ}, and we show that the energy at
unitarity depends universally on the effective range, as long as
Efimov states can be neglected.  Finally, in Sec.~\ref{sec:disc} we
compare variational approaches to the Bose polaron and discuss which
Hamiltonians and energy functionals can provide rigorous bounds on the
ground-state energy.


\section{Model}
\label{sec:model}

We consider a single impurity particle immersed in an interacting Bose
gas.  The combined system is described by the Hamiltonian
\begin{multline}
  \label{eq:Ham}
  H = \frac{\Hat{\vec p}^2}{2m_\text{I}}
  + \sum_i V_\text{IB}(\Hat{\vec x}_{\text{B},i}-\Hat{\vec x}) \\
  + \sum_i \frac{\Hat{\vec p}_{\text{B},i}^2}{2m_\text{B}}
  + \sum_{i<j} V_\text{BB}(\Hat{\vec x}_{\text{B},i} - \Hat{\vec x}_{\text{B},j}).
\end{multline}
Here, $\Hat{\vec p}$ and $\Hat{\vec x}$ denote the momentum and
position of the impurity of mass $m_\text{I}$, and
$\Hat{\vec p}_{\text{B},i}$ and $\Hat{\vec x}_{\text{B},i}$
characterize the bosons $i=1,\dotsc,N$ of mass $m_\text{B}$.  The
boson interaction $V_\text{BB}(\vec x)$ is assumed to be repulsive and
the impurity-boson interaction $V_\text{IB}(\vec x)$ is attractive.

The coupled system of impurity and bosons is conveniently analyzed in
the reference frame comoving with the impurity.  This is achieved by a
canonical transformation introduced by Lee, Low, and Pines
\cite{lee1953} and elaborated on by Girardeau \cite{girardeau1961}.
The transformation
$S=\exp(i\Hat{\vec x} \cdot \sum_i \Hat{\vec p}_{\text{B},i})$ leads
to the LLP Hamiltonian
\begin{multline}
  \label{eq:HLLP}
  H_\text{LLP} = S^{-1} H S
  = \frac{(\vec p_0-\sum_i\Hat{\vec p}_{\text{B},i})^2}{2m_\text{I}}
  + \sum_i V_\text{IB}(\Hat{\vec x}_{\text{B},i}) \\
  + \sum_i \frac{\Hat{\vec p}_{\text{B},i}^2}{2m_\text{B}}
  + \sum_{i<j} V_\text{BB}(\Hat{\vec x}_{\text{B},i} - \Hat{\vec x}_{\text{B},j}),
\end{multline}
in which the impurity operators have been eliminated, and $\vec p_0$
denotes the conserved total momentum of the system.  In the comoving
frame, the impurity potential $V_\text{IB}(\vec x)$ acts as a static
external potential centred at the origin, while the kinetic term
proportional to $\sim 1/m_\text{I}$ accounts for the recoil of the
impurity.  For bosons in the vicinity of the impurity this term leads
to induced interactions between bosonic particles in addition to their
inter-boson repulsion $V_\text{BB}(\vec x)$.

\textbf{{Homogeneous Bogoliubov Approximation.---}}In order to
appreciate the importance of the adequate inclusion of boson repulsion
we briefly review approximations to the model \eqref{eq:Ham} that are
frequently applied to the study of the Bose polaron problem. In the
formalism of second quantization Eq.~\eqref{eq:Ham} reads
\begin{align}\label{eq:Ham2Quant}
\hat H=&\sum_\vecp \frac{\vecp^2}{2m_I}\ded_\vecp\de_\vecp
+\sum_\vecp \frac{\vecp^2}{2m_B} \ad_\vecp\a_\vecp\nonumber\\
+&\frac{1}{\cal{V}}\sum_{
   \veck\veck'\vecq}V_\text{IB}(\vecq)
   \ded_{\veck'+\vecq}\de_{\veck'}\ad_{\veck-\vecq}\a_{\veck}\nonumber\\
+&\frac{1}{2\cal{V}}\sum_{\veck \veck'\vecq}
   V_{BB}(\vecq)\ad_{\veck'+\vecq}\ad_{\veck-\vecq}\a_{\veck}\a_{\veck'}.
\end{align}
Here $\mathcal{V}$ is the system volume, $\ded_\vecp$ and $\ad_\vecp$
are the creation operators of impurity and bosons, respectively, and
$V_\text{BB}(\vecq)$ and $V_\text{IB}(\vecq)$ are the Fourier
transforms of the respective interactions in Eq.~\eqref{eq:Ham}. Next,
the ladder operators of bosons are shifted using a simple canonical
coherent state transformation leading to
$\ad_\vecp\to \ad_\vecp +\delta_{\vecp,0} \sqrt{N_0}$.

This is then followed by the crucial \textit{Bogoliubov
  approximation}: All terms beyond quadratic order bosonic operators
are neglected, leading to the truncated Hamiltonian
\begin{align}\label{eq:HTrunc}
\hat H'=&\sum_\vecp \frac{\vecp^2}{2m_I}\ded_\vecp\de_\vecp
+\sum_\vecp \frac{\vecp^2}{2m_B} \ad_\vecp\a_\vecp\nonumber\\
+&\frac{N_0}{\mathcal{V}} V_{IB}(\mathbf{0}) +\frac{\sqrt{N_0}}{\cal{V}}\sum_{ \veck'\vecq}V_\text{IB}(\vecq)\ded_{\veck'+\vecq}\de_{\veck'}(\ad_{-\vecq}+\a_{\vecq})\nonumber\\
+&\frac{1}{\cal{V}}\sum_{ \veck\veck'\vecq}V_\text{IB}(\vecq)\ded_{\veck'+\vecq}\de_{\veck'}\ad_{\veck-\vecq}\a_{\veck}\\
+&\frac{g_\text{BB} N_0^2}{2\cal V}+\frac{g_\text{BB}N_0}{2\cal V}\sum_{\vecq\neq0} \left(2 \ad_\vecq \a_\vecq +\ad_\vecq\ad_{-\vecq}+\a_\vecq\a_{-\vecq}\right),\nonumber
\end{align}
where, for simplicity, we have chosen the example of a boson contact
interaction of strength $g_\text{BB}$. As a result of the Bogoliubov
approximation the bosonic part of the model can be diagonalized using
the standard Bogoliubov rotation
\begin{eqnarray}\label{BogTrafo}
\be_\vecp=u_\vecp \a_\vecp + v_{-\vecp}^* \ad_{-\vecp}\,\,,\,\,\bed_\vecp=u_\vecp^* \ad_\vecp + v_{-\vecp} \a_{-\vecp}.
\end{eqnarray}
While the Bogoliubov approximation thus allows to obtain a simple
dispersion relation for bosonic quasiparticles, it, however, captures
the effect of repulsion only within the Bogoliubov coefficients
$u_\vecp$ and $v_\vecp$. They give rise, e.g., to the modified
quasiparticle dispersion
$\omega_\vecp = \sqrt{\epsilon_\vecp(\epsilon_\vecp+ 2 n_0 g_{BB})}$
where, notably, the \textit{non-deformed}, homogenous boson density
$n_0$ appears.  It turns out that this approximate account of boson
repulsion is insufficient to self-stabilize the Bose polaron. Indeed,
the neglect of terms beyond quadratic order is responsible for the
apparent instability of the truncated Hamiltonian
\eqref{eq:HTrunc}\footnote{The instability becomes physical for a
  non-interacting BEC \cite{shchadilova2016quantum,
    drescher2021}.}.  Instead, it is crucial to keep all terms
beyond quadratic order which allows the repulsive interactions to act
as a stabilizing counter term to the impurity attraction. Including
these terms allows then to expand the theory not simply around the
homogenous BEC but around a BEC that is already deformed due to the
presence of the impurity (for a discussion of the one-dimensional case
see Ref.~\cite{jager2020}). As discussed in the following, this in
turn allows one to effectively map the strong coupling Bose polaron
problem onto a weakly interacting one.

\subsection{Gross-Pitaevskii functional}

Following this strategy we focus on a variational approach to the
ground state of the full Hamiltonian~\eqref{eq:HLLP}. We use a product
state \cite{pitaevskii2003}
\begin{align}
  \label{eq:Psi}
  \Psi(\vec x_1,\dotsc,\vec x_N) = \phi(\vec x_1) \dotsm \phi(\vec x_N)
\end{align}
where the condensate wave function $\phi(\vec x)$ is normalized to the
condensate particle number
$\int d^dx\, \abs{\phi(\vec x)}^2=N_0$\footnote{For weak boson
  interactions $N\approx N_0$.}.  Both $\phi(\vec x)$ and the
ground-state energy are found by minimizing the resulting variational
Gross-Pitaevskii (GP) energy functional
\begin{multline}
  \label{eq:gp}
  E_\text{GP}[\phi] = \frac{\bigl[\vec p_0-\int d^dx\, \Bar\phi (-i\nabla)\phi\bigr]^2}
  {2m_\text{I}} \\
  + \int d^dx \Bigl[ \frac{\abs{\nabla\phi}^2}{2m_\text{red}}
  + V_\text{IB}(\vec x)\abs{\phi(\vec x)}^2 + \frac g2
  \abs{\phi(\vec x)}^4 
  - \mu\abs{\phi(\vec x)}^2 \Bigr].
\end{multline}
Here $\mu$ is the chemical potential and we assume weak boson
repulsion represented in three dimensions by a contact interaction of
strength $g=4\pi a_\text{BB}/m_\text{B}$ with boson scattering length
$a_\text{BB}>0$.  Note that normal ordering of the impurity kinetic
term in Eq.~\eqref{eq:HLLP} contributes to the boson kinetic term with
reduced mass $m_\text{red}^{-1} = m_\text{B}^{-1} + m_\text{I}^{-1}$
\cite{shashi2014,shchadilova2016quantum,drescher2020}.

The energy functional Eq.~\eqref{eq:gp} exhibits two important
limiting cases: (i) for an infinitely heavy impurity
$m_\text{I}\to\infty$, the normal-ordered kinetic recoil term in the
first line of \eqref{eq:gp} vanishes and the standard GP energy
functional for bosons in a static external potential is recovered;
(ii) for a Bose polaron at rest ($\vec p_0=0$), and a radially
symmetric impurity potential $V_\text{IB}(\abs{\vec x})$, the wave
function $\phi(\vec x)$ is spherically symmetric and the recoil term
again vanishes ---it only re-appears beyond the product ansatz when
boson correlations are included \cite{drescher2020, Christianen1,
  Christianen2}.

We find the condensate wave function $\phi(\vec x)$ by minimizing the
GP functional \eqref{eq:gp} in the thermodynamic limit subject to the
boundary conditions $\abs{\phi(\vec x\to0)}<\infty$,
$\abs{\phi(\abs{\vec x}\to\infty)} = \sqrt{n_0}$ in terms of the
condensate density $n_0$ far away from the impurity.  For a radially
symmetric impurity potential $V_\text{IB}(\abs{\vec x})$ at rest
($\vec p_0=0$), the ground-state wave function is spherically
symmetric and real.  In the solution of the GP functional the energy
is universally expressed in units of the BEC bulk chemical potential
$\mu=gn_0$ and the distance from the impurity $r=\abs{\vec x}$ is
measured in units of the modified healing length
$\xi=1/\sqrt{2m_\text{red}\mu} = 1/\sqrt{8\pi(m_\text{red}/m_\text{B})
  a_\text{BB}n_0}$, which involves the \textit{reduced} mass and is
therefore larger than the usual bulk healing length
$\xi_0=1/\sqrt{2m_\text{B}\mu}$ of the BEC without impurity.

We define the polaron energy functional
$E[\phi] = E_\text{GP}[\phi] - E_\text{GP}[\phi_0]$ relative to the
energy of the unperturbed BEC with wave function $\phi_0=\sqrt{n_0}$.
For the three-dimensional case it is conveniently expressed in terms
of the scaled radial function $u(r) = r\phi(r)/\sqrt{n_0}$ as
\begin{multline}
  \label{eq:gpu}
  \frac{E[u]}\mu
  = 4\pi n_0 \int_0^\infty dr\,
  \Bigl[ \xi^2 \Bigl( \bigl(\frac{du}{dr}\bigr)^2 - 1\Bigr) +
    \frac{V_\text{IB}(r)}\mu u(r)^2 \\
    + \frac{(u(r)^2-r^2)^2}{2r^2} \Bigr].
\end{multline}
Here the boundary conditions for $\phi(\vec{x})$ translate to $u(0)=0$
and $u(r\to\infty)=r$. Numerically, $E[u]$ is minimized by global
optimization on $r\in[0,L]$ with $L=10\,\xi$ and an $r$-grid spacing
$\Delta r=0.05\,\xi$ much smaller than the potential range.
  
We thus obtain the scaling solution of the GP equation (GPE) in units
of $\mu$ and $\xi$ for a given dimensionless potential shape
$\tilde V(x=r/\xi)=V_\text{IB}(r)/\mu$.  This scaling solution is
\emph{universal} for arbitrary values of the condensate density $n_0$,
boson scattering length $a_\text{BB}$ and mass ratio
$m_\text{red}/m_\text{B}$ as long as GP theory applies
\cite{drescher2020, massignan2021}, with these parameters entering
only indirectly via $\mu$ and $\xi$.  For comparison with experiment,
and in order to visualize the effect of boson repulsion, the universal
solution can be rescaled to obtain the specific solution for desired
values of the boson scattering length $a_\text{BB}$ and the mass ratio
$m_\text{red}/m_\text{B}$ in density units of energy
$E_n=\hbar^2n_0^{2/3}/2m_\text{red}$ and length $n_0^{-1/3}$.

\subsection{Impurity potential}

For the impurity potential $V_\text{IB}(r)$ we consider two different
functional forms. This allows us to study the universality of the Bose
polaron by analyzing how quasiparticle properties depend on the
potential shape and range.  Specifically, we consider an attractive
Gaussian potential,
\begin{align}
  \label{eq:Vgauss}
  V_\text{gauss}(r) = -V_0 \exp[-(r/R)^2],
\end{align}
and an exponentially decaying potential,
\begin{align}
  \label{eq:Vexpon}
  V_\text{expon}(r) = -V_0 \exp[-r/R],
\end{align}
both of depth $V_0>0$ and range $R$. 

The low-energy scattering properties of the impurity and boson are
characterized by the impurity-boson scattering length $a_\text{IB}$
and the effective range $r_\text{eff}$, which determine the leading
terms of the effective range expansion of the momentum-dependent
scattering phase shift in three dimensions.  They are found by
numerically solving the Schr\"odinger equation for the scattering of a
boson with the impurity in the center-of-mass system via the potential
$V_\text{IB}(r)$. Equivalently, one may solve the first-order
nonlinear variable phase equation \cite{calogero1967}, which yields
\begin{align}
  \label{eq:var}
  a'(r) & = 2m_\text{red}V_\text{IB}(r) [r-a(r)]^2, \nonumber\\
  r_\text{eff}'(r) & = -4m_\text{red}V_\text{IB}(r) r^2\Bigl(\frac{r}{a(r)} -1 \Bigr)\nonumber\\
 &\quad\quad\quad \times\Bigl( \frac{r}{3a(r)} -1 + \frac{r_\text{eff}(r)}r \Bigr). 
\end{align}
Here $a(r)$ and $r_\text{eff}(r)$ obey the boundary conditions
$a(0)=0$, $r_\text{eff}(0)=0$, and account for the phase shift
accumulated by the scattering wave function (for the generalization of
the variable phase equation to singular potentials see
\cite{enss2020scattering}). Correspondingly, the differential
equations \eqref{eq:var} are integrated from $r=0\dotsc\infty$ which
yields the scattering length $a_\text{IB}=a(r\to\infty)$ and effective
range $r_\text{eff} = r_\text{eff}(r\to\infty)$.


\section{Effective potential}
\label{sec:effpot}

\begin{figure}[t]
  \centering
  \includegraphics[width=\linewidth]{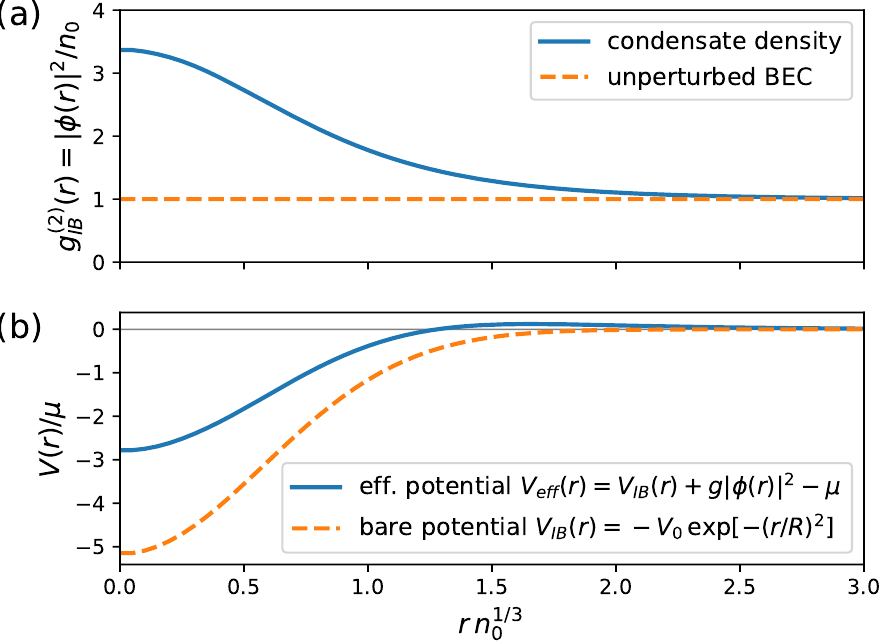}
  \caption{(a) Impurity-boson density-density correlation function
    $g_\text{IB}^{(2)}(r)$ as determined from the condensate wave
    function $\phi(r)$ as function of the distance $r$ from the
    impurity (blue solid). For an attractive impurity potential the
    wave function is enhanced near the impurity as compared to that of
    an unperturbed BEC $\phi_0(r)=\sqrt{n_0}$ (orange dashed).
    (b) The bare impurity potential (orange dashed) of Gaussian shape
    ($V_0/\mu=5.1651$, $R/\xi=0.81892$) has an effective range
    $r_\text{eff}=\xi$ and positive scattering length
    $a_\text{IB}=4\xi$, and correspondingly admits a bound state. An
    extra bosonic test particle is, however, subject to the effective
    potential (blue solid) that is weakened by the repulsion from the
    polaron cloud; while still attractive the effective potential is
    characterized by an effective, negative scattering length
    $a_\text{IB,eff}=-0.1\xi$ (and renormalized
    $\tilde r_\text{eff}=-207\xi$) which thus no longer supports a
    bound state. Hence, additional bosonic quantum fluctuations lead
    only to a weak, additional dressing of the impurity particle.}
  \label{fig:wavefct}
\end{figure}

First, we present results for the condensate profile around the
impurity that we find by minimizing the Gross-Pitaevskii energy
functional \eqref{eq:gp}--\eqref{eq:gpu}.  For attractive
impurity-boson interaction, the wave function $\phi(r)$ ---which, in
the comoving frame, directly yields the impurity-boson density-density
correlation function $g_\text{IB}^{(2)}(r)=|\phi(r)|^2/n_0$--- is
enhanced near the impurity, as shown in Fig.~\ref{fig:wavefct}(a).  In
this figure we have chosen a potential $V_\text{IB}(r)$ (dashed line
in Fig.~\ref{fig:wavefct}(b)) that is sufficiently deep to support a
two-body bound state at energy $-\eb<0$.  Correspondingly, the
potential is characterized by a positive impurity-boson scattering
length, in Fig.~\ref{fig:wavefct}(b) chosen as
$a_\text{IB} \approx 4\xi > 0$.

We thus realize a scenario as described in the introduction, which is
mimicked by a microtrap or a single-site Bose-Hubbard model. While at
the two-body level it suggests a dynamical instability where the
occupation of the bound state would grow without bounds, many-body
effects come to the rescue. Indeed, due to the repulsion between
bosons, each additional boson trying to participate in the formation
of the Bose polaron is subject to the effective potential
\begin{align}
  \label{eq:Veff}
  V_\text{eff}(r) = V_\text{IB}(r) + g\abs{\phi(r)}^2 - \mu,
\end{align}
that results from by the competition of the bare attractive potential
$V_\text{IB}(r)$ (dashed line in Fig.~\ref{fig:wavefct}(b)) and the
repulsion created by the already existing polaron cloud.

In essence, this effect can be understood as arising in an effective
density-functional theory (DFT) for the BEC particles in the presence
of the impurity: the bosons already attracted to the impurity screen
the attractive potential and make it shallower, as shown by the blue
line in Fig.~\ref{fig:wavefct}(b). Moreover, we find that they create
a small repulsive barrier at intermediate distance. As a consequence,
the \textit{effective} potential no longer admits a bound state and it
is correspondingly characterized by a \emph{negative} effective
impurity-boson scattering length $a_\text{IB,eff} \approx -0.1
\xi$. Hence, the single-particle excitation spectrum for each
additional boson is bounded from below: the dynamical instability is
replaced by \emph{Bose polarons self-stabilized by their dressing
  cloud}.


\section{Bose polaron energy and contact}
\label{sec:energy}

\begin{figure}[t]
  \centering
  \includegraphics[width=\linewidth]{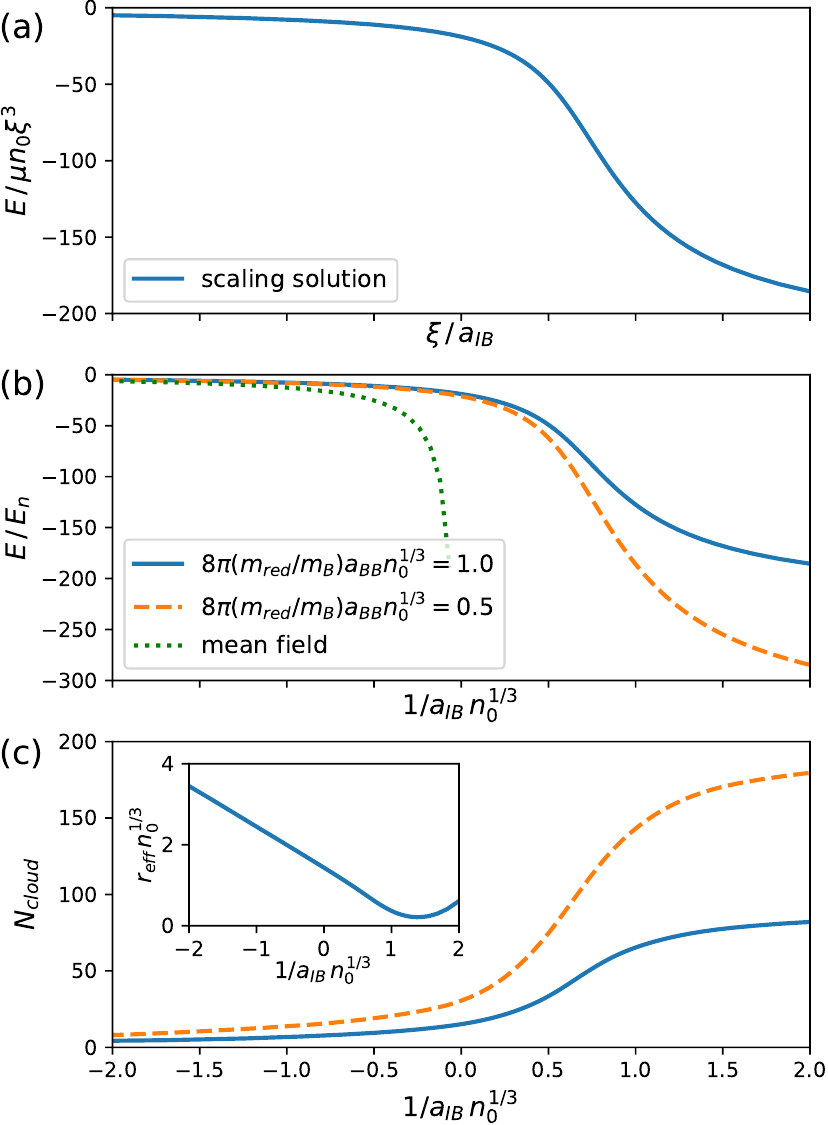}
  \caption{Bose polaron energy across an impurity-boson Feshbach
    resonance.  (a) Scaling solution for the polaron energy
    $E/\mu n_0\xi^3$ as function of the impurity-Bose interaction
    $\xi/a_\text{IB}$ for a Gaussian potential of fixed range $R=\xi$.
    (b) Polaron energy $E$ in density units
    $E_n=\hbar^2n_0^{2/3}/2m_\text{red}$ in dependence on the
    impurity-Bose interaction $1/a_\text{IB}n_0^{1/3}$ for different
    BEC gas parameters. The polaron binding energy is larger for weak
    boson repulsion
    $8\pi(m_\text{red}/m_\text{B}) a_\text{BB}n_0^{1/3}=0.5$ (orange
    dashed) compared to stronger repulsion
    $8\pi(m_\text{red}/m_\text{B}) a_\text{BB}n_0^{1/3}=1.0$ (blue
    solid); at weak coupling it approaches the mean-field result
    \eqref{eq:Emeanfield} (green dotted).  (c) Particle number
    $N_\text{cloud}$ within the polaron dressing cloud.  Inset: the
    effective range $r_\text{eff}$ for fixed potential range
    $Rn_0^{1/3}=1$ is smallest on the repulsive side.}
  \label{fig:Epol}
\end{figure}
    
The value of the Gross-Pitaevskii functional \eqref{eq:gp} evaluated
at the ground-state wave function determines the polaron energy
$E = E[\phi]$ relative to the homogeneous BEC.  The energy is shown in
Fig.~\ref{fig:Epol}(a) as a function of the dimensionless
impurity-boson interaction $\xi/a_\text{IB}$. It is always negative
for an attractive impurity potential. In Fig.~\ref{fig:Epol}(b) we
present the energy for a mobile impurity of arbitrary mass in units of
$E_n$ which depends on the BEC density $n_0$. Specifically, we show
results for two BEC gas parameters, which for equal mass of boson and
impurity, i.e., $m_\text{red}/m_\text{B}=1/2$, correspond to values
$n_0a_\text{BB}^3 = (4\pi)^{-3} = 5.0\times10^{-4}$ ($n_0\xi^3=1$) at
stronger boson repulsion and
$n_0a_\text{BB}^3 = (8\pi)^{-3} = 0.63\times10^{-4}$ ($n_0\xi^3=2.8$)
at weaker boson repulsion.  We find that, in absolute terms, the
polaron binding energy is larger for weaker boson repulsion where the
BEC can be more strongly deformed and thus acquire more attractive
potential energy.

At weak impurity-bath attraction $1/a_\text{IB}n_0^{1/3}\ll-1$ the
polaron energy approaches the mean-field value
\begin{align}
  \label{eq:Emeanfield}
  E_\text{mf} = \frac{2\pi a_\text{IB} n_0}{m_\text{red}}
  = 4\pi a_\text{IB} n_0^{1/3}\, E_n
\end{align}
irrespective of $a_\text{BB}$.  Within mean-field theory the polaron
energy diverges to $-\infty$ at unitarity
$1/a_\text{IB}=0$. Variational approaches based on the
\textit{trunctated} Hamiltonian Eq.~\eqref{eq:HTrunc} predict that the
inclusion of Bogoliubov corrections is not sufficient to heal this
instability, but rather results in a shift of the instability to the
repulsive side of the Feshbach resonance
\cite{shchadilova2016quantum}. In contrast, we find that going beyond
the Bogoliubov approximation by working with the full Hamiltonian
\eqref{eq:Ham} allows the boson repulsion to stabilize the polaron at
a finite ground-state energy that smoothly crosses over from the
attractive to the repulsive side of the Feshbach resonance.

The deformation of the BEC is also reflected in the number of bosons
participating in the formation of the polaron dressing cloud
\begin{align}
  \label{eq:Ncloud}
  N_\text{cloud} = 4\pi n_0 \int_0^\infty dr [u(r)^2-r^2].
\end{align}
As shown in Fig.~\ref{fig:Epol}(c), for a smaller gas parameter
(orange dashed line) the impurity attracts a larger polaron cloud
because the bosons are less repulsive.  Naturally, this larger
dressing cloud corresponds to the larger polaron binding energy found
in Fig.~\ref{fig:Epol}(b).

Our results in Fig.~\ref{fig:Epol} are shown for a constant range $R$
across the Feshbach resonance as applicable to experiments where the
microscopic range of interactions can typically not be tuned
synchronously with the scattering length. Thus the effective range
$r_\text{eff}$ varies in dependence on $a_\text{IB}$: The inset of
Fig.~\ref{fig:Epol}(c) shows the effective range that is obtained from
the scattering phase shift using Eq.~\eqref{eq:var}.  At constant
potential range $Rn_0^{1/3}=1$, the effective range $r_\text{eff}$
reaches a minimum on the repulsive side of the resonance
($a_\text{IB}>0$) and grows towards both weak-coupling limits
$a_\text{IB}\to0^\pm$.

\begin{figure}[t]
  \centering
  \includegraphics[width=\linewidth]{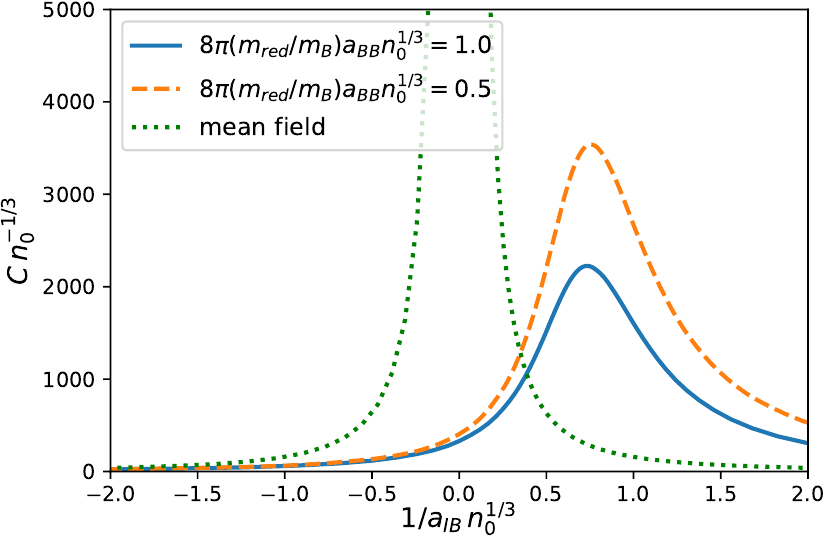}
  \caption{Tan contact $C$ of the Bose polaron for different boson
    repulsion (BEC gas parameter) across the impurity-boson Feshbach
    resonance.  The contact $Cn_0^{-1/3}$ obtained from
    Eq.~\eqref{eq:Contact} reaches its maximum on the repulsive side
    of the resonance and increases for weaker boson repulsion.  At
    weak coupling it approaches the mean-field result
    \eqref{eq:Cmeanfield} (green dotted).}
  \label{fig:contact}
\end{figure}

The variation of the polaron energy with $a_\text{IB}$ defines the
impurity contact parameter \cite{tan2008energetics, drescher2020,
  guenther2021, drescher2021, massignan2021} which characterizes
the impurity-boson correlations $g^{(2)}_{\text{IB}}(r)$ at short
distances outside the impurity potential:
\begin{align}
  \label{eq:Contact}
  C = \frac{8\pi m_\text{red}}{\hbar^2}\, \frac{\partial
  E}{\partial(-1/a_\text{IB})}
  = 4\pi n_0^{1/3}
  \frac{\partial(E/E_n)}{\partial(-1/a_\text{IB}n_0^{1/3})}.
\end{align}
The contact is shown in Fig.~\ref{fig:contact}: we find that it
reaches a maximal value on the repulsive side of the resonance.
Similar to the energy, we find that the contact is larger for smaller
boson repulsion $a_\text{BB}$ (blue solid line) where the BEC is more
strongly deformed and thus $g^{(2)}_{\text{IB}}(r)$ is enhanced (see
Fig.~\ref{fig:wavefct}).  At weak attractive interaction the contact
approaches the ground-state value of an impurity in an ideal BEC
($a_\text{BB}=0$) \cite{drescher2021}
\begin{align}
  \label{eq:Cmeanfield}
  C_\text{mf} = 16\pi^2 n_0 a_\text{IB}^2,
\end{align}
which  agrees with the mean-field result prediction.


\section{Universality}
\label{sec:univ}

\begin{figure}[t]
  \centering
  \includegraphics[width=\linewidth]{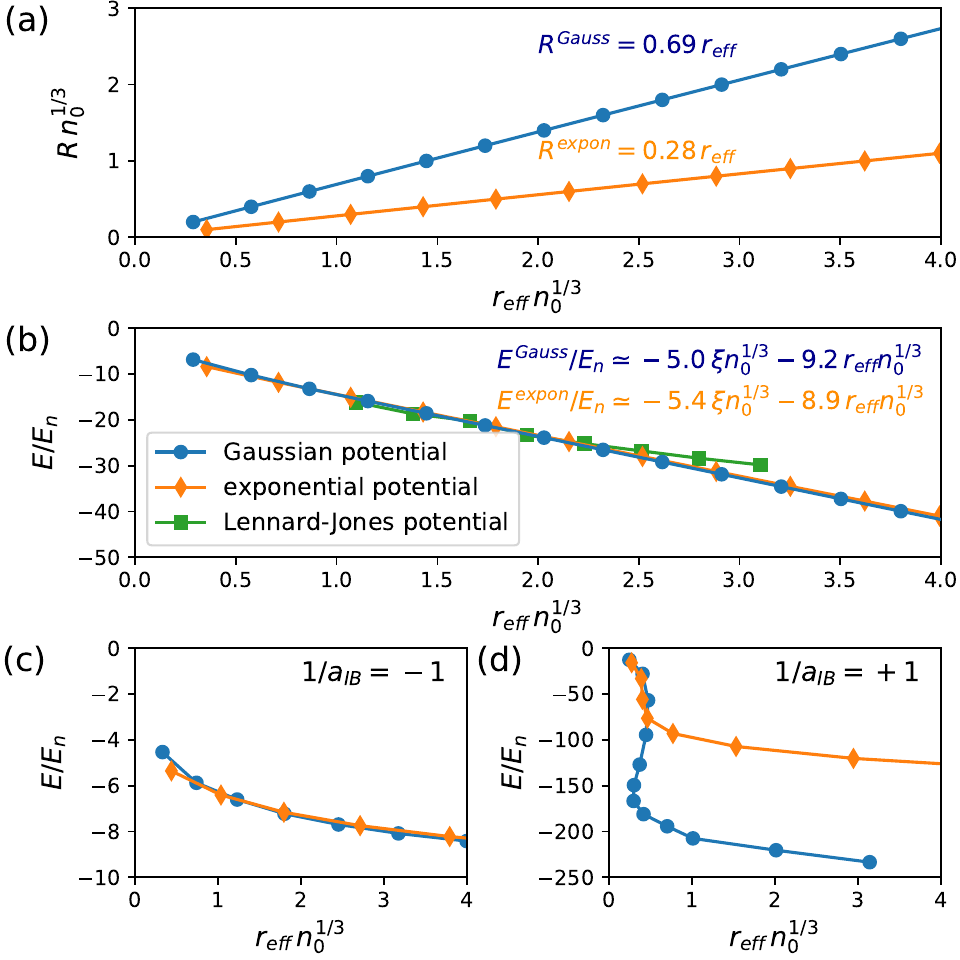}
  \caption{Universality of the Bose polaron. 
    (a) In order to yield the same effective range $r_\text{eff}$, the
    Gaussian and exponential potentials need to be tuned to different
    potential ranges $R$; here shown for fixed unitary scattering
    length $1/a_\text{IB}=0$.
    (b) Polaron energy $E$ as function of the effective range
    $r_\text{eff}$ at unitarity $1/a_\text{IB}=0$ for two different
    potential shapes.  The boson repulsion is set to
    $8\pi(m_\text{red}/m_\text{B}) a_\text{BB}n_0^{1/3}=1$.  The
    polaron energy coincides for both potential shapes and increases
    linearly with $r_\text{eff}$.  This universality also extends to
    Lennard-Jones potentials as shown by the green squares. 
    (c) The polaron energy at negative scattering length
    $1/a_\text{IB}n_0^{1/3}=-1$ is universal for $r_\text{eff}
    n_0^{1/3} \gtrsim 1$ for the parameters chosen.
    (d) Polaron energy at positive scattering length
    $1/a_\text{IB}n_0^{1/3}=1$.  For the Gaussian potential there is
    no unique polaron energy for intermediate values of $r_\text{eff}$
    (see text).}
  \label{fig:range}
\end{figure}

Finally, we test the notion of universality of the Bose polaron by
studying different shapes and ranges $R$ of the impurity potential.
Generally, we work in the regime where the potential range is much
larger than the boson scattering length, $R\gg a_\text{BB}$, where
local Gross-Pitaevskii theory has been shown to be applicable
\cite{chen2018}.  To test
universality in this regime, we specifically compare the predictions
following from the Gaussian potential $V_\text{gauss}(r)$ in
Eq.~\eqref{eq:Vgauss}, and the exponential potential
$V_\text{expon}(r)$ in Eq.~\eqref{eq:Vexpon} for various ranges $R$.

We find that when tuning the depth of the potentials to obtain equal
scattering length $a_\text{IB}=-\infty$ at equal range $R$, the
polaron energies are very different.  Instead, if $R$ is tuned to
yield the same effective range $r_\text{eff}$ for both potentials, as
shown in Fig.~\ref{fig:range}(a), remarkable agreement between the
polaron energies is found.  Indeed, as shown in
Fig.~\ref{fig:range}(b), for both potentials, the polaron energy
approximately follows a \emph{linear} scaling law for
$r_\text{eff}\gtrsim0.2\xi$,
\begin{align}
  \label{eq:range}
  \frac{E(1/a_\text{IB}=0)}{E_n}
  = -5.2(2)\,\xi n_0^{1/3} - 9.0(1)\, r_\text{eff}n_0^{1/3}.
\end{align}
Remarkably, this universality is found not only for the purely
attractive Gaussian and exponential potentials but extends also to
interactions featuring a repulsive contribution such as a
Lennard-Jones potential
$V_\text{LJ}(r)\sim\lambda R^{10}/r^{12}-R^4/r^6$, as illustrated in
Fig.~\ref{fig:range}(b).  This shows that in the Bose polaron problem
the momentum-dependent scattering phase shift is probed in the regime
where the effective range expansion is valid.  Our result complements
a recent GPE study which found a power-law scaling of the unitary
polaron energy at ranges shorter than the healing length,
$E/E_n \sim (r_\text{eff}/\xi)^{1/3}$ for
$a_\text{BB} \lesssim r_\text{eff} \ll \xi$ in the case of a
square-well potential \cite{massignan2021}.  The GPE approach can be
extended to even shorter ranges $r_\text{eff}\ll a_\text{BB}$ by using
a nonlocal generalization of Gross-Pitaevskii theory
\cite{drescher2020}.

Also on the attractive side of the resonance, see
Fig.~\ref{fig:range}(c) for $1/a_\text{IB}n_0^{1/3}=-1$, we find a
universal effective range dependence of the polaron energy, albeit not
a linear one.  Note that to cover the domain of effective ranges shown
in Fig.~\ref{fig:range}(d), for our parameters the Gaussian potential, unlike the
exponential, must be tuned over such depths that it supports more
than one bound state.  Consequently, there is no longer a unique mapping from
$r_\text{eff}$ to $E$.  This is illustrated in Fig.~\ref{fig:range}(d)
for the repulsive side, at $1/a_\text{IB}n_0^{1/3}=1$: universality
can at best hold in the vicinity of the first bound state, and only
for a limited interval of $r_\text{eff}$ values.  Furthermore, since
the Efimov effect modifies the polaron energy spectrum
\cite{zinner2013, levinsen2015, yoshida2018}, universality can also
depend on the three-body parameter
\cite{wang2012,SchmidtZwerger2012,langmack2018, field2020}.


\section{Discussion}
\label{sec:disc}

The variational principle provides a powerful tool to compute both
ground-state and dynamical properties of quantum many-body systems.
However, in order to make full use of its predictive power it is
essential to understand the limitations of approximations applied to
the Hamiltonian to be analyzed. In this regard the strong-coupling
Bose polaron is a case in point. The full Bose polaron Hamiltonian
\eqref{eq:Ham} ---and equivalently Eq.~\eqref{eq:Ham2Quant}--- is
bounded from below for repulsive boson interaction, and hence
variational wave functions give rigorous bounds on the ground-state
energy.  For finite Bose repulsion, the polaron energy remains finite
for any Bose-impurity scattering length, including resonant
interactions, and the ground state represents a strong-coupling Bose
polaron that is self-stabilized by its own dressing cloud.

In contrast, when applying the Bogoliubov approximation to the full
model \eqref{eq:Ham} by truncating terms of higher-than quadratic
order in the boson operators, the resulting, truncated Hamiltonian
$\hat H'$ in Eq.~\eqref{eq:HTrunc} becomes unbound from
below. Crucially, this results in an instability of the Bose polaron
problem that is solely an artefact of this approximation.

Quite remarkably, the instability of the truncated Hamiltonian
$\hat H'$ becomes, however, only evident when considering
wavefunctions that account for more than two phonon excitations from
the homogenous BEC. For instance, for a Chevy-type wavefunction
\cite{chevy2006,rath2013,li2014} that itself is truncated at the
single excitation level, the terms beyond quadratic order in the
repulsive interactions have vanishing expectation value.  Thus,
incidentally, the Chevy ansatz yields \textit{the same} prediction
when applied to both the full and the truncated model. Thus due to its
tremendous simplicity the Chevy ansatz becomes immune to the
instability of approximate Hamiltonian $\hat H'$.
 
However, while being a well-defined approach, the simple Chevy ansatz
misses the fact that at the weak Boson repulsion (as typically present
in cold gases) the polaron cloud---even within the full model
Eq.~\eqref{eq:Ham}---can contain an exceedingly large number of bosons
(Fig.~\ref{fig:Epol}(b)).  Such a large local deformation of the BEC
is naturally captured by the inhomogeneous wave function
\eqref{eq:Psi}. Crucially, while accounting for an arbitrary number of
boson excitations, when applied to the full Hamiltonian \eqref{eq:Ham}
it still leads to a bounded energy functional \eqref{eq:gp}. Its
solution shows that the smaller the boson repulsion and the wider the
impurity potential, the larger the polaron cloud becomes. The
product state approach is complementary to the Chevy ansatz
including its extensions to multi-boson excitations
\cite{levinsen2015, shi2018, yoshida2018}, and it becomes particularly
accurate for soft impurity potentials where it is justified to ignore
bosonic correlations. Remarkably, the case of extremely soft
potentials is realized with Rydberg excitation immersed in BECs. In
this case it was predicted that up to hundreds of atoms can be bound
to the single impurity leading to the creation of Rydberg polarons
\cite{schmidt2016}. Since for Rydberg impurities the range of
interactions $R$ dramatically exceeds the interparticle distance, our
local GP theory applies and provides a so far missing explanation as
to why the experimental observation of Rydberg polarons
\cite{Camargo2018} is described exceptionally well by a coherent state
approach \cite{schmidt2018}.

Recently also first steps to the understanding of the complementary,
intermediate regime of short-range impurity potentials ---with yet
large dressing clouds--- has been achieved by using an extension to
nonlocal Gross-Pitaevskii theory \cite{drescher2020}.  In conjunction
with our present result, these combined new approaches resolve a
fundamental shortcoming of the Bogoliubov approximation: while the
formulation of the interacting Bose gas in terms of Bogoliubov
quasiparticles is exact, the additional approximation to neglect the
residual interaction between phonons is not.  In particular, the
quadratic Bogoliubov mean-field Hamiltonian is unbounded from below
for the strong-coupling Bose polaron, most obviously in the regime
where a two-body bound state appears on the repulsive side of the
resonance \cite{shchadilova2016quantum,grusdt2017}.

As discussed above, the Chevy-type ansatz applied to the truncated the
Bogoliubov Hamiltonian \cite{li2014} still yields a finite polaron
energy since it is of such low order in boson excitations that it is
not sensitive to the truncated part of the Hamiltonian.  However, when
the coherent state ansatz or higher-order excitation extensions of the
Chevy ansatz are applied to the truncated
Hamiltonian \eqref{eq:HTrunc} they can lead to divergencies in the
ground-state energy.  In the case of the coherent state ansatz
\cite{shchadilova2016quantum, drescher2019} the divergence is due to
the large occupation of excitations in the vicinity of the impurity
not counteracted by boson repulsion.  Instead the
local extension to the truncated Bogoliubov approach studied in the
present work (see also \cite{field2020, guenther2021, massignan2021})
as well as its nonlocal counterpart \cite{drescher2020} provides a
stable starting point for strong-coupling Bose polaron dynamics, and
we showed how it can find an effective description in terms of a
renormalization of the impurity-boson potential
(Fig.~\ref{fig:wavefct}).

Beyond our treatment of the two-particle impurity-boson correlations,
three-body and higher-order correlations give rise to the Efimov
effect and three-body recombination.  The Efimov effect can occur
either between one impurity and two bosons \cite{levinsen2015,
  shi2018, yoshida2018} or between two impurities and one boson
\cite{zinner2013, naidon2018}.  These few-body effects can be captured
by Gaussian variational wave functions
\cite{Christianen1,Christianen2}. Alternatively, extensions of the Chevy ansatz
to two or more independent bosonic excitations \cite{levinsen2015} can
be employed. The latter approach was applied in the analysis of the
truncated model \eqref{eq:HTrunc} and universal scaling depending on
the three-body parameter was found \cite{shi2018,yoshida2018}.  It
remains an interesting open question how universal three-body physics
carries over to the many-body case in a dense bosonic medium when the
full Hamiltonian Eq.~\eqref{eq:Ham} is considered \cite{levinsen2021}.

Finally, in ultracold atomic gases the boson repulsion originates from
an attractive van-der-Waals potential between atoms. This results in
the existence of deeply bound molecular states into which atoms can
decay in three-body recombination.  These deeply bound states are
neither accounted for in variational approaches nor in quantum Monte
Carlo \cite{penaardila2015}.  Up to now most experiments probe the
metastable Bose polaron state of matter on transient time scales where
the deeply bound states ---arising both from the fundamentally
attractive Bose-Bose and Bose-impurity potentials--- can be ignored,
and thus the stable variational approach discussed in our work is well
applicable. However, as one starts to explore longer time scales or
the build-up of more complex correlated states of impurities,
understanding the impact of the fundamental dissipative nature arising
from deeply bound states becomes essential and requires the
development of new theoretical approaches to quantum impurity
problems.

\begin{acknowledgments}
  This work is supported by the Deutsche Forschungsgemeinschaft (DFG,
  German Research Foundation), project-ID 273811115 (SFB1225 ISOQUANT)
  and under Germany's Excellence Strategy EXC2181/1-390900948 (the
  Heidelberg STRUCTURES Excellence Cluster). R.~S. is supported by the
  Deutsche Forschungsgemeinschaft (DFG, German Research Foundation)
  under Germany's Excellence Strategy -- EXC-2111 -- 390814868
  (Excellence Cluster `Munich Center for Quantum Science and
  Technology').
\end{acknowledgments}

\bibliography{all}

\end{document}